\documentstyle[pre,aps,tighten,epsf,floats]{revtex}
\begin{document}
\renewcommand
\baselinestretch{2}
\large
\title{\LARGE Efficiency and current reversals in spatially inhomogeneous ratchets \vspace{0.5in}}
\author{\large Debasis Dan,$^{1,*}$ Mangal C. Mahato,$^{2}$ and
A. M. Jayannavar$^{1,\dagger }$ \vspace{0.2in}}
\address{\large $^1$ Institute of Physics, Sachivalaya Marg,
Bhubaneswar 751005, India}
\address{\large $^2$ Department of Physics, Guru Ghasidas University,
 Bilaspur 495009, India}
\maketitle
\begin{abstract}
\vspace{1.0in}
\large

         Efficiency of generation of net unidirectional current in an 
adiabatically driven symmetric periodic potential system is studied. 
 The efficiency shows a maximum, in the case of an
 inhomogeneous system with spatially varying periodic friction coefficient, as
a function of temperature. The ratchet is not most efficient when
it gives maximum current. The direction of current may also be
reversed as a function of noise strength when, instead, an asymmetric 
periodic potential is considered.
\end{abstract}
\newpage


              Motion of a particle in a medium generally involves
frictional resistance. The frictional resistance or the friction
coefficient  influences the dynamical properties, such as the rate of 
relaxation of the system, but does not affect the equilibrium 
properties . In some inhomogeneous
systems the friction coefficient could be space dependent and such
systems are not rare in nature. The motion of particles in such
inhomogeneous system is difficult to describe theoretically because
the effect of the space dependent friction cannot be incorporated as
due to an effective potential. The motion of a Brownian particle in an 
inhomogeneous system with periodically varying space-dependent
friction coefficient is of much recent interest \cite{But,Kam,Mill,Jay,Par,Zhao,Bao,dan1,dan2,Mah,Kam2,takn} .

             Recently it has been shown that mobility of  over-damped
Brownian particles in washboard potential exhibit 
stochastic resonance in inhomogeneous systems without the 
application of periodic input signal \cite{dan1,dan2}, i.e, the 
probability current shows a maximum as function of temperature.
The mobility also exhibits a resonance like phenomenon as a function
of external field strength ( constant in time) and noise induced slowing
down of particle in an appropriate parameter regime \cite{dan2}. These effects 
could not have been possible in over-damped  homogeneous systems with 
uniform friction coefficient. Also, it has been shown that in these 
inhomogeneous systems it is
possible to obtain net unidirectional current in periodic 
potential systems in a thermal bath but subjected 
to an external zero mean time-periodic field or random forces \cite{Mill,Jay,Par,Bao}. 
 The efficiency of such rocked thermal ratchet system to achieve
unidirectional current has been examined. For this we have followed the
 method of stochastic energetics \cite{seki,tak1,tak2}.
 Earlier it was claimed that thermal
fluctuations monotonically reduce the efficiency of an adiabatically
rocked thermal ratchet \cite{tak1}. In the present work we mainly explore the
possibility of obtaining a current  in 
an inhomogeneous medium with varying space-dependent friction
coefficient when quasi-statically rocked by a zero-time-averaged
external field. Also, the dependence of the efficiency of operation of 
such systems on thermal fluctuations present in the system is
examined. In these systems we show that thermal fluctuations
facilitate energy conversion by ratchet systems ,
contrary to earlier claims \cite{tak1}. Moreover the direction of current may
also be reversed as a function of noise strength in the presence of
asymmetric periodic potential in the same quasi-static limit. 

         As usual, the efficiency of a machine( system) is defined as
the ratio of useful work accomplished by the system to the total
energy expended on the system to get the work done. We study the
motion of an over-damped Brownian particle in a potential $V(q)$ subjected 
to a space dependent friction coefficient $\gamma (q)$ and an external
force field $F(t)$ at temperature $T$. The motion is described by
the Langevin equation \cite{Par,dan1,dan2,Mah,Sancho}
\begin{equation}
  \label{Langv}    
  \frac{dq}{dt} = - \frac{(V'(q)-F(t))}{\gamma (q)} - k_{B}T
  \frac{\gamma '(q)}{[\gamma (q)]^{2}} + \sqrt{\frac{k_{B}T}{\gamma
      (q)}} \xi (t) ,
\end{equation}
where $\xi (t)$ is a randomly fluctuating Gaussian white noise with zero
mean and correlation :\\  $ <\xi (t) \xi (t')> = 2 \delta (t-t')$. We take
$V(q) = V_{0}(q) + qL$, where $V_{0}(q + 2n\pi) = V_{0}(q) = -sin(q)$,
$n$ being any natural integer. $L$ is a constant force ( load)
representing the slope of the washboard potential against which the
work is done. Also, we take the friction coefficient $\gamma (q)$ to
be periodic :\\ $\gamma (q) = \gamma _{0}(1 - \lambda \sin (q + \phi))$, 
where $\phi$ is the phase difference with respect to $V_{0}(q)$. The
equation of motion is equivalently given by the Fokker-Planck equation
\begin{equation}
  \label{Fok_Plk}      
  \frac{\partial P(q,t)}{\partial t} = \frac {\partial}{\partial q} \frac{1}{\gamma
    (q)} [k_{B}T \frac {\partial P(q,t)}{\partial q} + (V'(q) - F(t)
  )P(q,t)] . 
\end{equation}
This equation can be solved for the probability current $j$ when $F(t)
= F_{0}$ = constant, and is given by \cite{dan1,dan2,Ris}
\begin{equation}
  \label{curr}      
  j = \frac{k_{B}T ( 1 - \exp(-2 \pi(F_{0} - L)/k_{B}T))}{\int _{0}^{2
      \pi}\exp(\frac{-V_{0}(y) + (F_{0} - L)y}{k_{B}T}) dy
    \int_{y}^{y+2\pi}\gamma(x)\exp \frac{V_{0}(x) - (F_{0} 
      - L)x)}{k_{B}T}} dx .
\end{equation}
It may be noted that even for $L = 0,\; j(F_{0})$ may not be equal to
$-j(-F_{0})$  for $\phi \neq 0, \pi$. This fact leads to the
rectification of current (or unidirectional current ) in the presence
of an applied a.c field $F(t)$. We assume $F(t)$ changes slowly enough, 
i.e, its frequency is smaller than any other frequency related to
relaxation rate in the problem. For a field $F(t)$ of a square wave
amplitude $F_{0}$, an average current over the period of oscillation
is given by, $<j>\; =\; \frac{1}{2}\,[j(F_{0}) + j(-F_{0})]$. This 
particle current can even flow against the applied load $L$ and
thereby store energy in useful form. In the quasi-static limit
following the method of stochastic energetics it can be shown \cite{tak1,tak2} 
that the input energy $R$ \\ ( per unit time) and the work $W$ ( per unit time) 
that the ratchet system extracts from the external noise are given by 
$R = \frac{1}{2} F_{0}[j(F_{0})-j(-F_{0})]$ and $W =
\frac{1}{2}L[j(F_{0})+j(-F_{0})]$ respectively. Thus the efficiency 
( $\eta$) of the system to transform the external fluctuation 
to useful work is given by 
\begin{equation}
  \label{effi}         
  \eta = \frac{L[j(F_{0}) + j(-F_{0})]}{F_{0}[j(F_{0}) - j(-F_{0})]}.
\end{equation}

           Earlier, it has been found that for homogeneous systems,
$\gamma(q) = \gamma_{0}, \eta$ monotonically decreases as $T$ is
increased from zero in the quasi-static limit \cite{tak1}. We find, 
however, that in the case of
inhomogeneous systems, when $\gamma(q)$ is periodic, it is possible to 
maximize $\eta$ (see below) as a function of temperature by suitably
choosing the parameters $\phi, \lambda$ and $F_{0}$ for given $L$.

\begin{figure}
  \centerline{\epsfysize=19cm \epsfbox{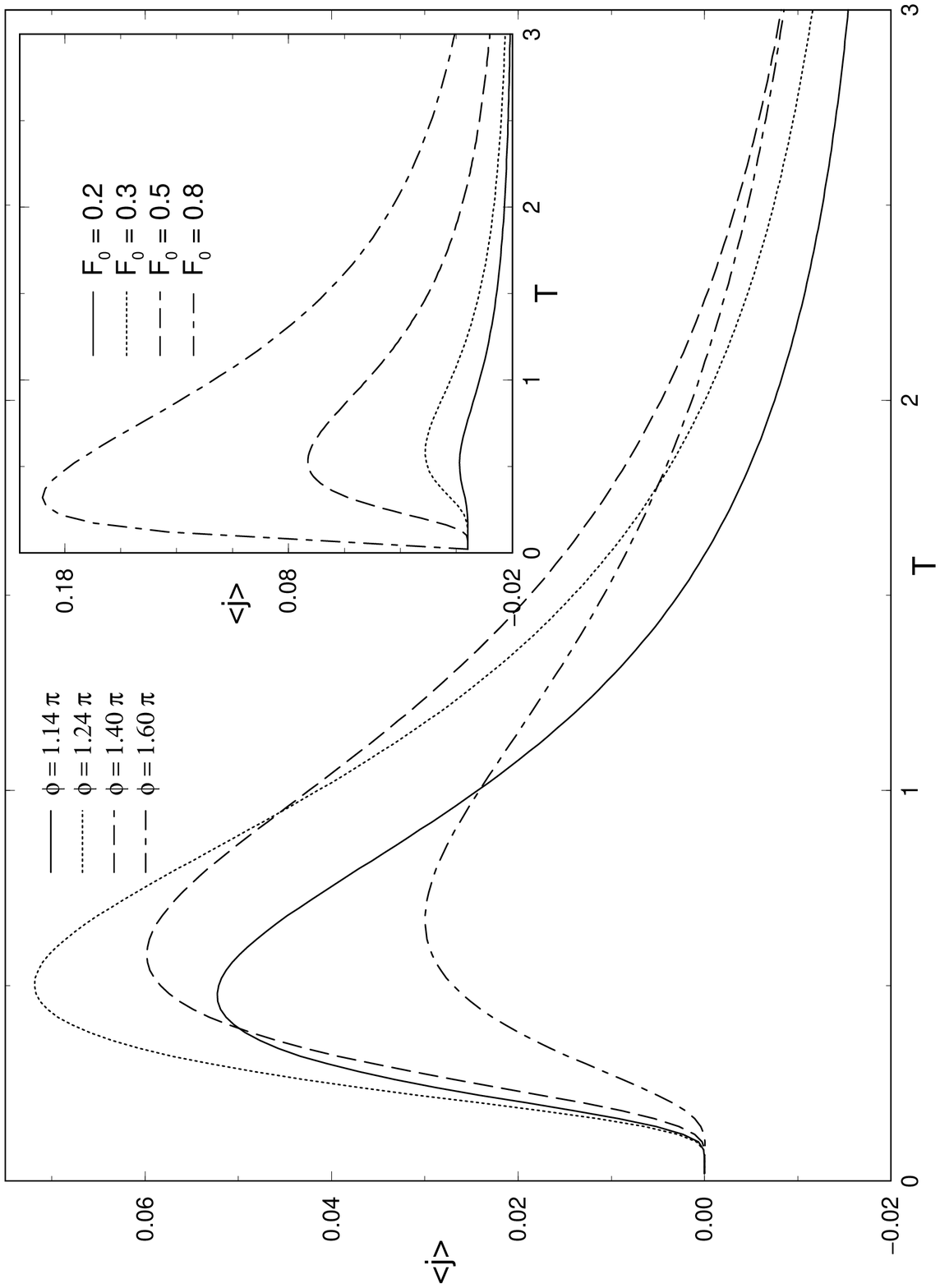}}
  \vspace{1.0in}
  \caption{}  \label{compare_curr}
\end{figure}

\begin{figure}
  \centerline{\epsfysize=17cm \epsfbox{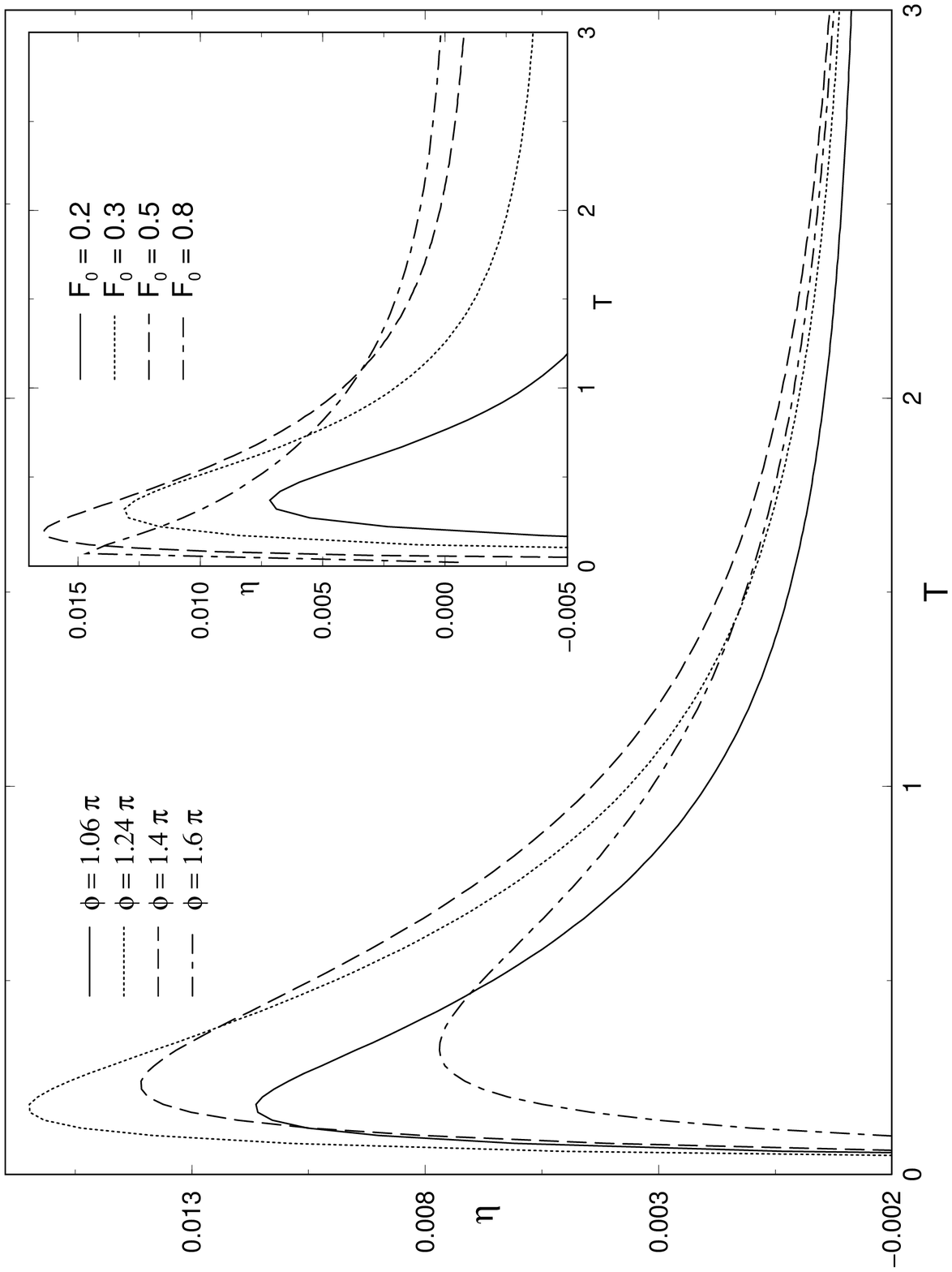}}
  \vspace{1.0in}
  \caption{}   \label{compare_effi}
\end{figure}

\begin{figure}
  \centerline{\epsfysize=19cm \epsfbox{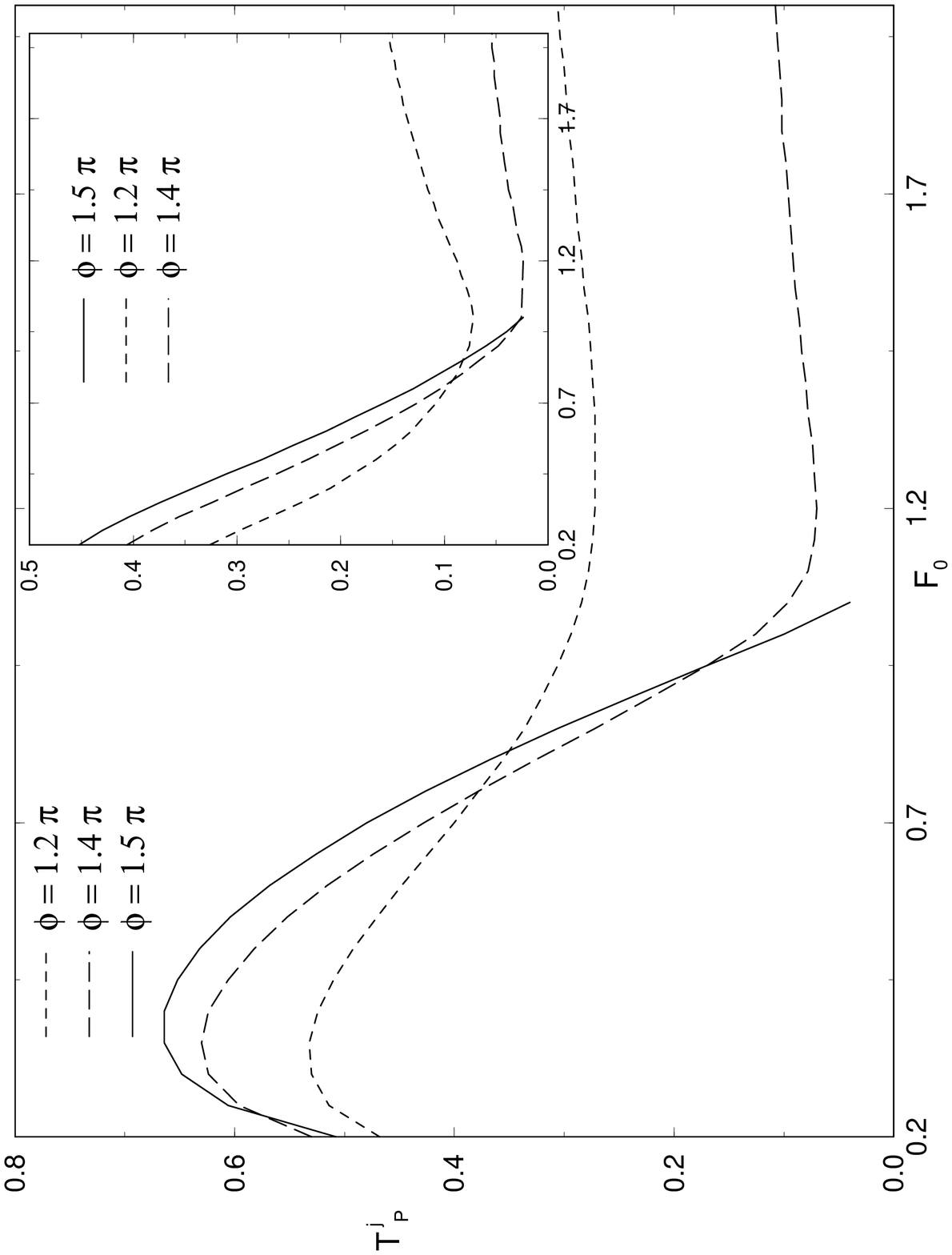}}
  \vspace{1.0in}
  \caption{}   \label{T_p}
\end{figure}

          It is found, numerically, that the effect of $\lambda$, the
amplitude of periodic modulation of $\gamma(q)$, for instance on $<j>$
is more pronounced for larger value of $\lambda (0 \leq \lambda <
1)$. We choose $\lambda = 0.9$ throughout our work. Contrary to the
homogeneous friction case where the net current goes to zero , the 
detailed analysis of the Eqn. (\ref{curr}) shows that the net
current $<j>$ quickly saturates monotonically to a value equal to
$\frac{-\lambda}{2} \sin \phi$ as a function of $F_{0}$ irrespective of 
the value of temperature $T$. All the interesting features are
captured at smaller values of $F_{0}$ itself. We shall confine
ourselves to $F_{0} < 1$, the threshold value at which the barrier of
the potential $V_{0}(q) = -\sin q$ just disappears. $\phi$ determines 
the direction of asymmetry of the ratchet. For $\phi > \pi$, we have 
a forward moving ratchet ( current flowing in the positive direction) 
and for $\phi < \pi$ we have the opposite, in the presence of external 
quasi-static force $F(t)$. When the system is homogeneous
($\lambda = 0$), $<j> = 0$ for $L = 0$ and for all values of $F_{0}$ and $T$, 
because the potential $V(q)$ is symmetric. However, when $\lambda \neq 
0, <j> \neq 0$ for all $\phi \neq 0, \pi$. 
Fig.(\ref{compare_curr}) clearly
illustrates this. In fig.(\ref{compare_curr}) we have plotted $<j>$
(in dimensionless units) for various values of $F_{0}$ at
$\phi = 1.3 \pi$ as a function of temperature $T$ (in dimensionless
units). Henceforth all our variables like $ <j>, F_{0}, T$ are in 
dimensionless units \cite{Ris}.
The inset shows the behaviour of $<j>$ with temperature for 
various $\phi$ when $F_{0} = 0.5$. We have chosen $L=0.02$ (corresponding to
a positive mean slope of $V(q)$) and $F_{0} = 0.3$. One would expect,
for $\lambda = 0$, $<j> \: < 0$ for all values of $F_{0}$ and $T \neq
0$. Fig.(\ref{compare_curr}) shows that $ <j> \: > 0$ for all values
of $\phi$ ( the net current being up against the mean slope of
$V(q)$). Though the
result is counterintuitive, but it is still understandable. For
$j(-F_{0})$ the particle is expected to have acquired higher mean
velocity before it hits the large $\gamma(q)$ region and hence faces
maximum resistive force and therefore though $j(-F_{0})$ is negative
it is small in magnitude compared to $j(F_{0})( >0)$, where the
particle faces just the reverse situation. The average current $< j >
$ exhibits maximum as a function of temperature. At large value of
temperature, $< j >$ becomes negative. In this regime of high
temperature the ratchet effect disappears and the current which is
negative is in response to the load.
The peaking behaviour of $<j>$ as a function of $T$ is due to the 
synergetic effect of the thermal
fluctuations and the space dependent friction coefficient $\gamma(q)$. Of
course, this result is obtained in the quasi-static limit when the time 
scale of variation of $F(t)$ is much larger compared to any other
relevant time scales involved in the system. Fig.(\ref{compare_effi}) 
shows the efficiency $\eta$ of flow of current against the load (
slope) $L$. The parameter values chosen for
figs.(\ref{compare_curr} and \ref{compare_effi}) are same.  
\emph{$\eta$ shows a maximum as a function of
  temperature}. It is to be noted that the temperature corresponding
to maximum efficiency is not close to the temperature at which the
current $<j>$ is maximum. From this we conclude that 
\emph{the condition of maximum current  does not correspond to maximum 
  efficiency of current generation.} This fact has been pointed out 
earlier but in a
different system \cite{tak2}. Fig.(\ref{T_p}) shows how the temperature $T =
T_{P}^{j}$ at which $<j>$ peaks vary vis-a-vis the temperature $T =
T_{P}^{\eta}$ ( inset) at which $\eta$ peaks as a function of $F_{0}$ for 
various values of $\phi$ as shown in the figure. We find that
$T_{P}^{\eta} \neq T_{P}^{<j>}$ for any value of $F_{0}$ and $\phi$. 

\begin{figure}
  \centerline{\epsfysize=19cm \epsfbox{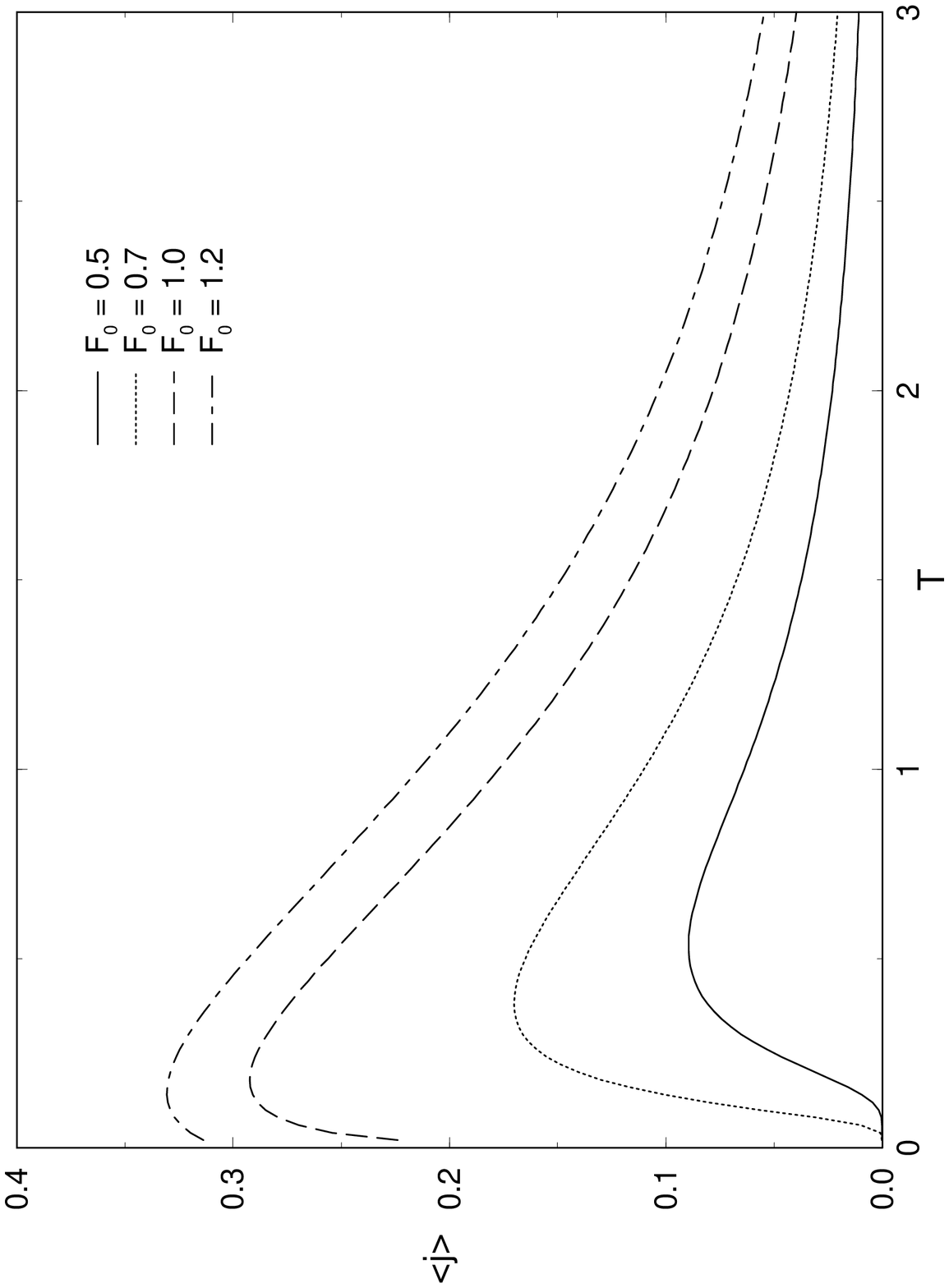}}
  \vspace{1.0in}
  \caption{}  \label{0_load}
\end{figure}

\begin{figure}
  \centerline{\epsfysize=19cm \epsfbox{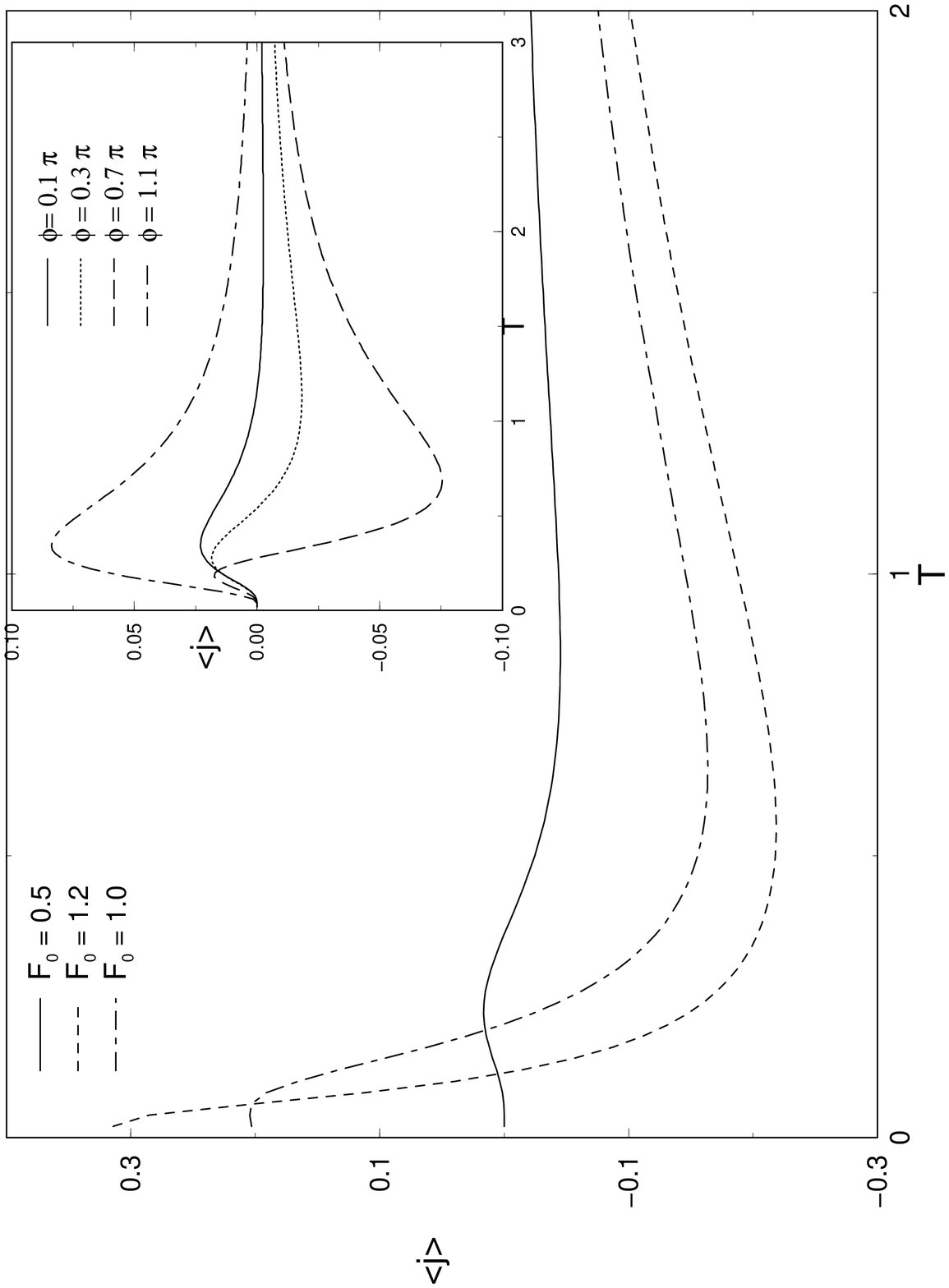}}
  \vspace{1.0in}
  \caption{}  \label{curr_rev}
\end{figure}

       In fig.(\ref{0_load}) we have plotted the average current $< j
>$ in the absence of load as function of temperature for various
values of $F_{0}$ for given $\phi = 1.3 \pi$. It shows that in the absence of
load $L$, $<j>$ never changes sign from positive to negative. It is
possible, however, to get a current reversal as a function of
temperature if, instead of a symmetric potential one 
considers an asymmetric potential $V_{0}(q) = -\sin q -\frac{\mu}{4}\sin
2q$ ( where $\mu$ lies between 0 and 1, describes an asymmetry
parameter) \cite{Zhao}. In fig.(\ref{curr_rev}) we have plotted $< j
>$ in the absence of load as function of temperature with asymmetry
parameter $\mu = 1 $.  In fig.(\ref{curr_rev}) we notice that the 
current reverses its direction as function of temperature or noise strength. 
This current reversal with temperature in the
adiabatic limit is possible only when the system is inhomogeneous. 
In homogeneous systems ( but asymmetric potential) current
reversal is possible when the frequency of the field swept $F(t)$ is
large \cite{Hanggi}.  The initial rise in the value of $<j>$
in fig. (\ref{curr_rev}), in the asymmetric potential case, can mainly be
attributed to the existence of two separate threshold values of $F_{0}$ 
for flow of current in the two directions. Here the current reversals
arise due to the combined effect of $\phi$ and $\mu$. It can be noted
that in the parameter range that we have considered the direction of
current for $\phi = 0$ and $\mu \neq 0$ is opposite to that when $\phi
\neq 0$ and $\mu = 0$. 

            We, thus, conclude that it is possible to achieve
reasonable efficiency in obtaining net unidirectional current in an
adiabatically rocked ratchet in an inhomogeneous system with space
dependent friction coefficient by suitably tuning the parameter values 
of the system. The efficiency shows a maximum as a function of
temperature as does the net current $<j>$, though they do not occur at 
the same temperature. In our case thermal fluctuations facilitate the energy
conversion. Also current reversal is possible in
inhomogeneous systems even in an adiabatically rocked but asymmetric
potential ratchet system.

\section*{Acknowledgements} 
            MCM acknowledges partial financial support and hospitality 
from the Institute of Physics, Bhubaneswar. MCM and AMJ acknowledge
partial financial support from the Board of Research in Nuclear
Sciences, DAE, India.            

\hspace{1.8in} \Large REFERENCES \hfill
  
\newpage
\hspace{1.5in} \large  FIGURE CAPTIONS .\vspace{0.5in}
\large

Fig. 1. The net current $<j>$ as a function of $T$ for various values
of $\phi$ at $F_{0} = 0.5$. The inset shows the variation of $<j>$
with $T$ for different $F_{0}$ at $\phi = 1.3 \pi$. $L = 0.02$ for both 
the figures.

\vspace{.3cm}
Fig. 2. Efficiency $\eta$ as a function of $T$ for various values of
$\phi$ at $F_{0} = 0.5$. The inset shows the variation of $\eta$ as a
function of $T$ for different values of $F_{0}$ at $\phi = 1.3
\pi$. $L = 0.02$ for both the figures.

\vspace{.3cm}
Fig. 3. Temperature $T_{p}^{j}$, corresponding to peak current as a
function of $F_{0}$ for various values of $\phi$.  
The inset shows the
variation of temperature $T_{p}^{\eta}$, corresponding to peak
efficiency as a function of $F_{0}$ for various values of $\phi$. $L = 
0.02$ in this case

\vspace{.3cm}
Fig. 4. Current $<j>$ as a function of $T$ for various values 
of $F$ at $\phi = 1.3 \pi$. $L = 0$ in this case.

\vspace{.3cm}
Fig. 5. Current $<j>$ in asymmetric potential for various values of 
$F$ (for diferent values of $\phi$ in the inset) at $\phi = 1.3 \pi$( at $F =
0.5$ in the inset ) when load $L =0 $.

\end{document}